\documentclass{article}

\usepackage{amsfonts}
\usepackage{cite}
\usepackage{graphicx}
\usepackage{dcolumn}
\usepackage{bm}
\usepackage{amsmath}
\usepackage{amssymb}
\usepackage{hhline}
\usepackage[normalem]{ulem}
\usepackage{authblk}

\usepackage{color}

\usepackage{hyperref}
\hypersetup{colorlinks=true,citecolor={blue},linkcolor={blue},urlcolor={blue}}

\newcommand{\clr}[1]{{ #1}}

\date{}
\begin{document}

\title{\textbf{Steady state oscillations of circular currents in concentric polariton condensates}}

\author{
{\Large Vladimir~Lukoshkin$^{1,2}$, 
Evgeny Sedov$^{3,2,4,*}$,
Vladimir~Kalevich$^{1,2}$,
Z.~Hatzopoulos$^{5}$,
P.~G.~Savvidis$^{6,7,5,8}$,
Alexey Kavokin$^{6,7,2,9}$ }\\
$^1${Ioffe Institute, Russian Academy of Sciences, 26 Politechnicheskaya, 194021 St-Petersburg, Russia}\\
$^2${Spin Optics Laboratory, St. Petersburg State University, Ulyanovskaya 1, St. Petersburg 198504, Russia}\\
$^3${Russian Quantum Center, 100 Novaya Street, Skolkovo, Moscow 143025, Russia}\\
$^4${Stoletov Vladimir State University, Gorky str. 87, Vladimir 600000, Russia}\\
$^5${FORTH-IESL, P.O. Box 1527, 71110 Heraklion, Crete, Greece}\\
$^6${Key Laboratory for Quantum Materials of Zhejiang Province, School of Science, Westlake University, 18 Shilongshan Rd, Hangzhou 310024, Zhejiang, China}\\
$^7${Institute of Natural Sciences, Westlake Institute for Advanced Study, 18 Shilongshan Road, Hangzhou, Zhejiang Province 310024, China}\\
$^8${Department of Materials Science and Technology, University of Crete, P.O. Box 2208, 71003 Heraklion, Crete, Greece}\\
$^9${Moscow Institute of Physics and Technology, Institutskiy per., 9, Dolgoprudnyi, Moscow Region, 141701, Russia}\\
{*evgeny\_sedov@mail.ru}
}

\maketitle
\begin{abstract}
Concentric ring exciton polariton condensates emerging under non-resonant laser pump in an annular trapping potential support persistent circular currents of polaritons.
The trapping potential is formed by a cylindrical micropillar etched in a semiconductor microcavity with embedded quantum wells and a repulsive cloud of optically excited excitons under the pump spot.
The symmetry of the potential is subject to external control via manipulation by its pump-induced component.
In the manuscript, we demonstrate excitation of concentric ring polariton current states with predetermined vorticity which we trace using interferometry measurements with a spherical reference wave. 
We also observe the polariton condensate dynamically changing its vorticity during observation, which results in pairs of fork-like dislocations on the time-averaged interferogram coexisting with azimuthally homogeneous photoluminescence distribution in the micropillar. 
\end{abstract}

\section*{Introduction}

The most successful approach to the fine control of light is by coupling it with matter.
The brightest representatives of such coupled matter-light systems are exciton-polariton condensates.
They are usually formed in specially designed layered semiconductor heterostructures, optical microcavities, containing embedded quantum wells (QWs)~\cite{kavokinBook2017}.
Quantized microcavity photons strongly couple with excitons in QWs and form quasiparticles exciton-polaritons.
The latter, being bosons, are able to form mesoscopic coherent states of exciton-polariton condensates~\cite{ScienceScience199,NatPhys10803}. 

Due to the finite lifetime of polaritons, polariton condensates possess a non-equilibrium nature.
They can exist for a long time only at the balance of gain and losses in the presence of external pumping.
Engineering the gain-loss landscape along with the shape of the confining potential in the microcavity plane enables to design the vector field of internal fluxes of exciton polaritons within the condensate cloud~\cite{PhysRevX5021025,PhysRevA98053812,PhysRevB85155320, PhysRevB102201114,PhysRevResearch3013099}.
The most advantageous shape of the potential landscape for disclosing internal flux behavior of polariton condensate is annulus. 
In the annular geometry, the polariton condensate demonstrate properties of a superfluid liquid~\cite{NatPhys6527,RevModPhys85299} supporting persistent circular polariton currents and votrices~\cite{PhysRevB97195149,ACSPhoton71163,PhysRevResearch3013072}, \clr{including those in the presence of external and effective magnetic fields~\cite{SciRep1122382,PhysRevB101104308}.}
Ring polariton condensates are characterized by discrete winding numbers which allow to consider them as macroscopic quantum objects, potential opening ground for applications in quantum information processing~\cite{NatPhysRev4435}.

In this context, an interesting and important object of study in annular potentials is concentric ring condensates, which so far have been deprived of attention in comparison with single ring condensates.
\clr{The concentric ring geometry is of a significant interest as compared to single ring condensates.
It enables one study experimentally correlations between superfluid currents of exciton polaritons in a geometry that preserves the azimuthal symmetry.
Such correlations are necessarily present if both rings are located in the same trap, however, their origin is far from obvious.
One can speculate that the direction of propagation of currents in both rings must coincide in a stationary case.
Our experiments show that the real picture is much more rich and complex. These studies pave the way towards the realization of double-qubit gates based on exciton-polariton condensates as discussed in Ref.~\cite{NatPhysRev4435}.}
In Ref.~\cite{PhysRevB103075305} the polariton condensates in concentric ring trapping potentials have been studied theoretically.
The coflowing currents of polaritons (cowinding vortices) in the condensates linked with each other by a Josephson junction have been predicted.
The authors of Ref.~\cite{PhysRevB104165305} claim that rotation of polariton ring condensates may be characterized by different winding numbers, and it must be accompanied by formation of a Josephson vortex in the annular junction.
The introduction of a long Josephson junction is an elegant way to describe splitting of the condensate into two concentric rings.
This, however, introduces undesirable perturbations into the internal kinetics of polaritons, making important the transversal (radial) kinetics of polariton flows~\cite{PhysRevB104165305}. 

In our manuscript, we present an experimental study of concentric ring polariton condensates in a single ring trapping potential.
Such potentials can be created in a microcavity sample at the stage of its manufacturing by etching of pillars~\cite{PhysRevB100245304,YaoZBpaper2022} or by non-resonant optical pumping~\cite{PNAS1118770,PhysRevB92035305}.
The optically induced potential is formed by a cloud of incoherent excitons emerging in a sample under the laser pump beam of a given shape.
Due to the repulsive interaction, the cloud acts as a barrier for polaritons.
The reservoir replenished by optical pumping, in addition, acts as a continuous source of polaritons for the condensate state.
The obvious advantage of the optical potentials is the controllability of their shape during the experiments. 
In addition, due to the optical gain, the optical potentials contribute to the stable occupation of the excited states~\cite{PhysRevB103115309}.  
Polaritons occupy the mode characterized by the fastest-growing rate  under the competition of losses and gain~\cite{PhysRevB103115309}.  
Herewith the gain of the mode is strongly governed by its overlap with the exciton reservoir, which is subject to control by the pump power as well as by the shape and position of the pump spot~\cite{PhysRevB97045303}.
\clr{In the pot-like geometry, the radius of the trap for polaritons is an effective control parameter for the condensate-reservoir overlapping.
The excitation of the higher excited radial and azimuthal condensate modes in the trap with the increasing radius has been discussed in~Refs.~\cite{PhysRevB103115309,PhysRevB107045302}.
In the ring geometry, the key parameter is the width of the trap ring.
In Ref.~\cite{SedovNanosystPChM13608} one has demonstrated that with an increase in the width of the ring (the diameter of the outer wall of the trap), the lower in the radial quantum number~$n$ states become lossy, and the condensate has to occupy the higher in~$n$ states.}

Here we combine the two approaches to the trapping of polaritons.
We consider formation of the condensates in a cylindrical micropillar cavity under the  non-resonant optical pump beam focused close to the center of the pillar.
The localizing potential is formed by the edges of the pillar and by the repulsive reservoir of optically induced excitons under the pump spot.
\clr{The diameter of the pillar ($\text{30} \, \mu \text{m}$) was chosen such that the condensate occupied the first excited radial mode with~$n=1$.}

Formation of petal double concentric ring condensates in a fully optical trap has been reported in Ref.~\cite{PNAS1118770}.
In the experiment of geometry similar to one considered in this manuscript, some of us have observed two and multiple concentric ring condensates in cylindrical micropillars of different diameters~\cite{PhysRevB91045305}. 
However, the azimuthal dynamics of polaritons was not studied.
In this manuscript, we report on the experimental observation of double concentric ring polariton condensates with persistent azimuthal polariton currents.
\clr{We want to emphasize that despite the double concentric ring shape, the considered  polariton state represents a single polariton condensate described by a single wave function which is a first excited radial mode of the ring-shaped potential trap.
This differs it from two adjacent concentric ring condensates separated by an annular Josephson junction, see, e.~g.~Ref.~\cite{PhysRevB104165305}.}
We get access to the phase properties of the condensate via measurements of interference of photoluminescence (PL) of the condensates with the reference spherical wave.
We demonstrate the control over the PL helisity  of the concentric ring condensates. 
We also show the regime of oscillations of the polariton condensate helisity.
In measurements, it is manifested in a homogeneous in azimuthal direction distribution of the condensate density together with an even number of fork-like dislocations on interferometry images.
We support our observations with analysis based on the generalized Gross-Pitaevskii equation.

\section*{Results}

Concentric ring polariton condenastes are excited by a non-resonant pump beam from cw laser normal to the microcavity plane and focused close to the center of the pillar.
In each experiment, we perform two types of measurements.
We measure the near-field PL distribution of the polariton condensate and the interference of light emitted by the condensate with the reference spherical wave.
For the interferometry measurements we use the Mach-Zehnder interferometer.
We obtain the reference spherical wave by magnifying a small peripheral area of the condensate image with use of a convex lens.
More details on the experimental setup and sample details can be found in~Refs.~\cite{PhysRevB97195149,ACSPhoton71163}.

\begin{figure}[tb!]
\begin{center}
\includegraphics[width=0.5\linewidth]{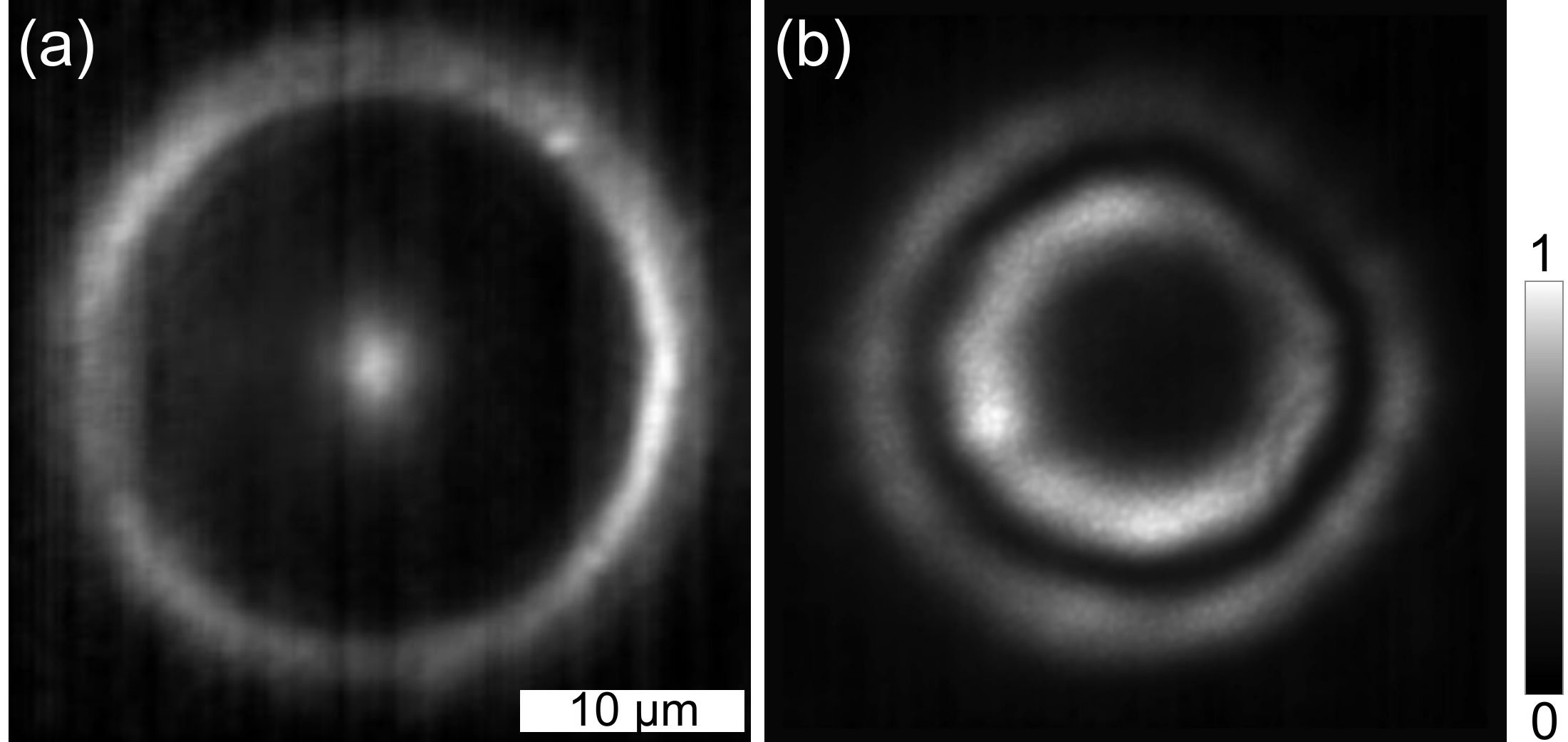}
\end{center}
\caption{ \label{FIG_EXP_Rings}
\clr{Real-space images of PL} of the sample under the non-resonant optical pump below (a) and above (b) the polariton condensation threshold.
\clr{In (a) the central peak and the outer ring are the luminescence from the laser-induced exciton reservoir and from the edge of the pillar due to scattering, respectively.
In (b) the double ring intensity pattern corresponds to PL from the mesoscopically occupied state of the exciton polariton condensate.}
}
\end{figure}

\begin{figure}[tb!]
\begin{center}
\includegraphics[width=\linewidth]{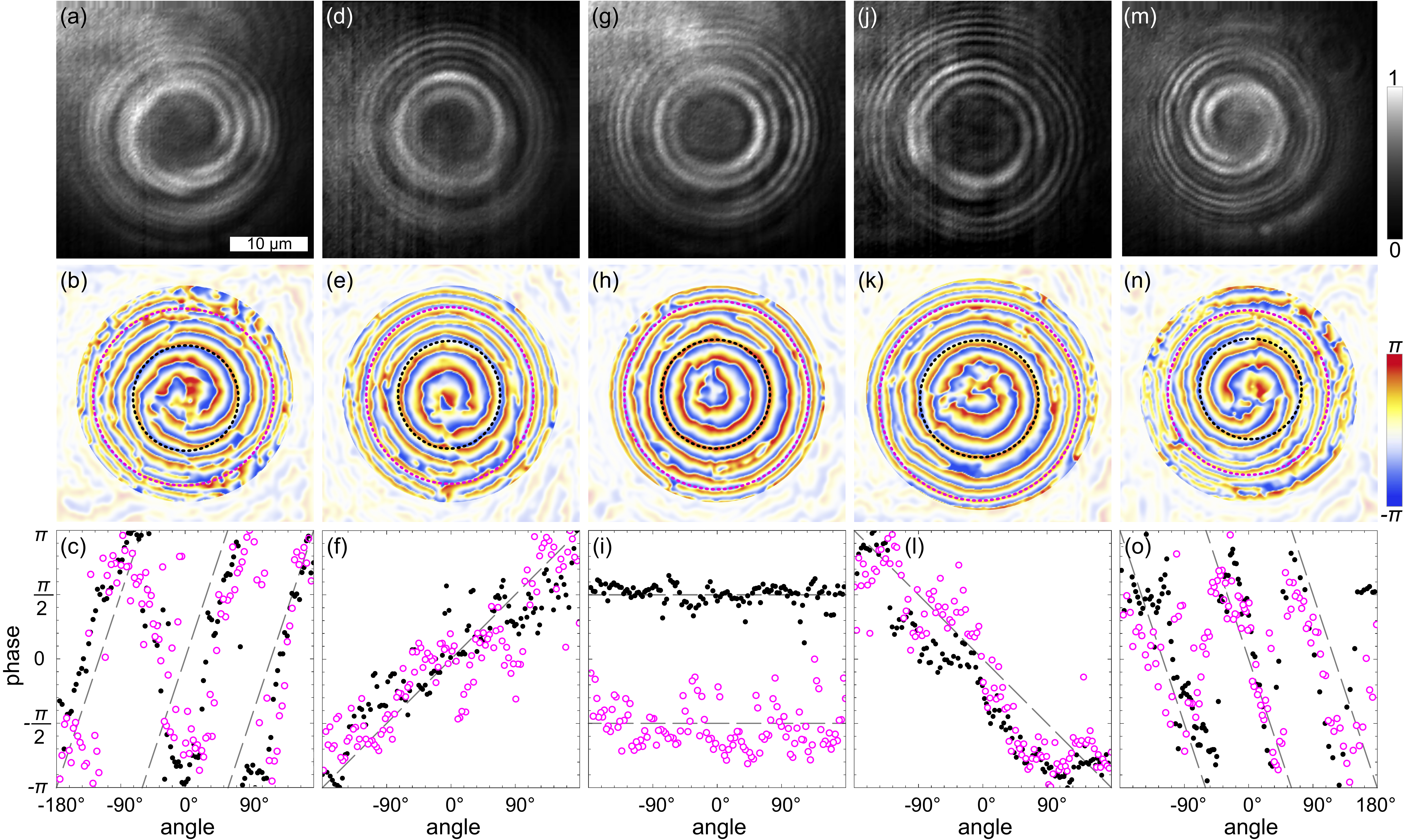}
\end{center}
\caption{ \label{FIG_EXP_Single_and_Triple}
Observation of concentric ring polariton condensates that either contain (a--f, j--o) or do not contain (g--i) azimuthal currents.
\clr{Top panels show interferometry images obtained from the interference of PL of the condensates in the pillar microcavity and the spherical reference wave in the Mach–Zehnder interferometer. 
Middle panels show the phase of the condensates relative to the phase of the reference wave extracted from the corresponding interferometry images.
Bottom panels show the phase variation along the maxima of the condensate rings (indicated by black and magenta dashed curves in middle panels).}
In bottom panels, the dashed lines are guides for the eye corresponding to the linear variation of the phase $\varphi$ with the azimuthal angle $\theta$, $\varphi = m\theta + const.$ with $|m|=0,  1$ and $3$.
\clr{The coordinate origin $0^{\circ}$ was chosen arbitrary to minimise the number of breaks of the dependence $\varphi(\theta)$.}
}
\end{figure}

\subsection*{Concentric polariton vortices}

At the pump power below the condensation threshold, $P < P_{\text{th}}$, occupation of high wave vector polariton states prevails due to the bottleneck effect~\cite{APL92042119,PhysRevB65205310,SciRep75542}.
Luminescence from the sample in the below-threshold pumping regime is shown in Fig.~\ref{FIG_EXP_Rings}(a).
It enables visualizing the shape of the effective trapping potential for polaritons.
The central luminous peak corresponds to the potential hill of the reservoir of optoinduced excitons under the pump spot.
The  peripheral ring illuminates the edge of the pillar due to scattering. 

At the pump power above the polariton lasing threshold, polaritons condense to a macroscopic quantum state of the polariton condensate, occupying the bottom of the annular potential trap. 
The condensate is characterized by a complex wave function $\Psi (\mathbf{r}) = \sqrt{\rho (\mathbf{r})} \exp [i \varphi (\mathbf{r})] $, where $\rho (\mathbf{r})$ and $\varphi (\mathbf{r})$ are the density and phase distribution of the condensate in the pillar plane.
Photoluminescence of the condensate, containing information of the polariton density distribution, obtained at the pump power of $P \approx 1.5 P_{\text{th}}$ is shown in Fig.~\ref{FIG_EXP_Rings}(b).
Such double concentric ring shape is typical for all polariton condensates discussed below.

The vector field of the polariton current density in a general case is given by $\mathbf{J} (\mathbf{r}) = \text{Im} [\Psi ^* (\mathbf{r}) \nabla \Psi (\mathbf{r})] = \rho (\mathbf{r}) \nabla \varphi (\mathbf{r})$.
If the azimuthal distribution of the polariton condensate density is homogeneous, the azimuthal component of the current density up to a constant is determined by  $J_{\theta} \propto \partial _{\theta} \varphi$, so we can measure the azimuthal currents of polaritons in the condensate by measuring variation of the condensate phase over the micropillar.
\clr{To quantify the circulation of the polariton flux, it is convenient to introduce the orbital angular momentum (OAM) per particle $\ell = N^{-1} \int _{-\infty} ^{\infty} [x J_y(\mathbf{r}) - y J_x(\mathbf{r})] d \mathbf{r}$ and the winding number
$m = (2 \pi)^{-1} \oint _{\mathcal{P} } d \varphi$, which is the azimuthal quantum number, a topological charge of the circulating state characterizing the variation of the phase along the closed path~$\mathcal{P}$~\cite{PhysRevResearch3013072,PhysRevB104165305}.
$N =\int _{-\infty} ^{\infty} |\Psi(\mathbf{r})|^2 d \mathbf{r}$ is the occupancy of the condensate state.
The orbital angular momentum $\ell$, being the indicator of internal azimuthal currents in the polariton condensate, is the continuous variable that takes into account peculiarities of both the density and phase distribution of the condensate.
At the same time, the winding number~$m$, being a measure of vorticity (the velocity of the phase variation around the vortex core), is an integer~\cite{PhysRevResearch3013072}. 
For azimuthally homogeneous polariton condensates, values of OAM $\ell$ and the winding number $m$ coincide.
}

To reveal the condensate phase distribution, $\varphi (\mathbf{r})$, we measure the interference of the condensate with the coherent reference spherical wave.
The upper panels in Fig.~\ref{FIG_EXP_Single_and_Triple} show the interferometry images of several polariton condesates with different winding numbers, $m = 0, \pm 1$ and~$\pm 3$.
The interference fringes on the images corresponding to current states (with  $m \ne 0$) represent spirals with the direction of twist determined by the sign of $m$ and with the number of arms determined by~$|m|$.
For the no-current state (Fig.~\ref{FIG_EXP_Single_and_Triple}(g--i)), the interference fringes represent a set of concentric rings.
In all panels, two annular regions are clearly distinguishable (by higher brightness), corresponding to the position of the spatial components of the condensates.
To excite polariton current states with different winding numbers $m$ we  weakly (by a submicrometer distance) displaced the pump spot from the center of the pillar, reducing the symmetry of the problem and perturbing the condensate.
The double ring shape of the condensates, however, had not been destroyed by such subtle displacement. 
The obtained condensates with persistent azimuthal currents were highly stable and were observed until the pumping was turned off.
\clr{After turning off and then turning on the pumping, the phase of the condensate retained the direction of its swirling.
This indicates that the perturbations introduced by displacing the pump spot can make the system chiral and allow one to choose both the absolute value and the sign of OAM of the condensate in an arbitrary way.}

Using the Fourier-transform method adapted for closed-fringe patterns~\cite{PhysRevB97195149}, from each interferogram in Fig.~\ref{FIG_EXP_Single_and_Triple} we extract the phase of the condensate relative to the phase of the reference spherical wave, see middle row panels in~Fig.~\ref{FIG_EXP_Single_and_Triple}.
The phase distribution repeats the spiral shape of the interference fringes.
For convenience of interpretation, in the lower panels of Fig.~\ref{FIG_EXP_Single_and_Triple} we show the variation of the phase along the dashed circles, $\varphi(\theta)$, in the middle panels corresponding to the position of the ridges of the condensate rings. 
The black and magenta dots correspond to the inner and outer rings, respectively.
One can see that for all observed condensates, the azimuthal phase variation is similar for the inner and outer polariton condensate rings, and both rings are characterized by equal winding numbers~$m$.

\subsection*{Concentric ring polariton condensates with oscillating vorticity}

\begin{figure}[tb!]
\begin{center}
\includegraphics[width=\linewidth]{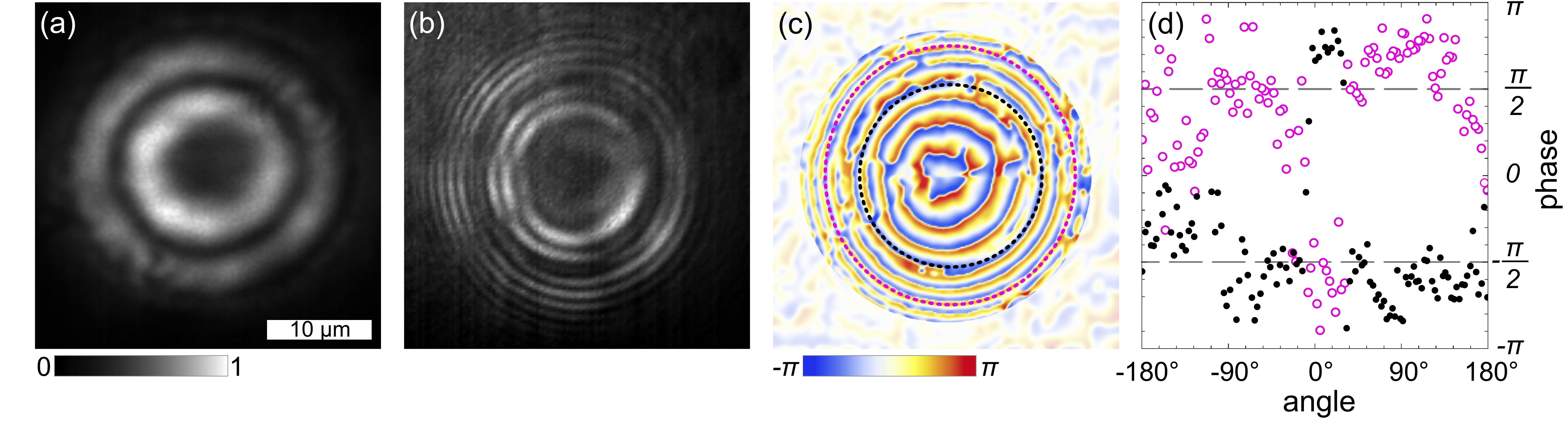}
\end{center}
\caption{ \label{FIG_EXP_Superp}
Observation of the concentric ring polariton condensate with oscillating vorticity.
(a) \clr{Real-space image of time-averaged PL} of the condensate, (b) the interferometry image \clr{obtained from the interference of PL of the condensate and the spherical reference wave}, (c) the  \clr{phase of the condensate relative to the phase of the reference wave} extracted from panel~(b), and (d) the phase variation around the condensate ring \clr{ridges indicated by black and magenta dashed circles in panel~(c).
Dashed lines are guides for the eye indicating constant phases~$\pm \pi/2$.}
}
\end{figure}

\begin{figure}[tb!]
\begin{center}
\includegraphics[width=.9\linewidth]{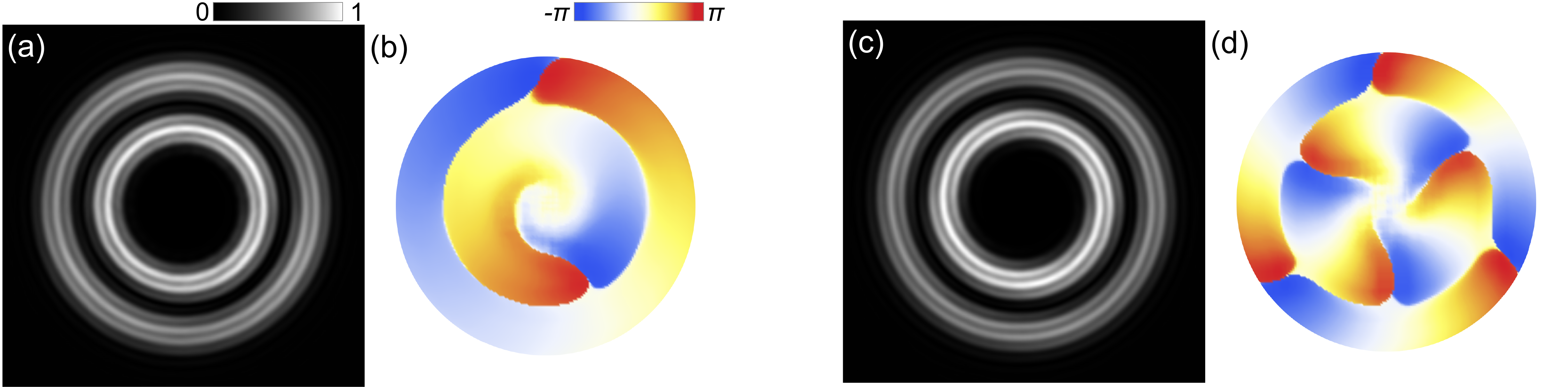}
\end{center}
\caption{ \label{FIG_SIMUL_Rings}
Simulated concentric ring polariton condensates with azimuthal currents.
Panels (a,c) show interferometry images \clr{obtained by simulating interference of the polariton condensates with the spherical reference wave.
Panels (b,d) show the corresponding spatial distribution of phases of the simulated  condensates.}
Topological charge is $m=1$ for (a,b) and $m=3$ for~(c,d).
\clr{Details of the model and values of the parameters used for simulations are given in Methods.}
}
\end{figure}

For a polariton condensate with the homogeneous density distribution  in the azimuthal direction, in view of the continuity of the condensate wave function and its derivative, we can reasonably expect that its phase monotonically changes along the condensate rings.
Figure~\ref{FIG_EXP_Superp} shows an example of the experimental observations, that contradicts these expectations.
Smooth concentric condensate density rings (a) coexist with the concentric ring  interferometry fringes with two breaks each (b) at around 8 and 9:30 o'clock positions.
One should mention that in different experiments with similar interferometer settings, the position of the breaks remained approximately the same.
On the restored phase distribution in Fig.~\ref{FIG_EXP_Superp}(c), the phase jumps take place at the position of the breaks.
The phase jumps on the ridges of the condensate rings by $\pm \pi$ are more clearly seen in~Fig.~\ref{FIG_EXP_Superp}(d). 
In our previous studies~\cite{ACSPhoton71163,PhysRevResearch3013072}, we have demonstrated that in a steady state polariton condensate the phase jump accompanies the density dip to ensure single-valuedness of the condensate wave function. 
Since no any noticeable density perturbations are observed in Fig.~\ref{FIG_EXP_Superp}(a), we assume that we observe not a stationary polariton condensate state, but the time-averaged density and interferometry images of the condensate, whose winding number changes in time.
In the Discussion section below we give our arguments for this assumption.

\begin{figure}[tb!]
\begin{center}
\includegraphics[width=0.7\linewidth]{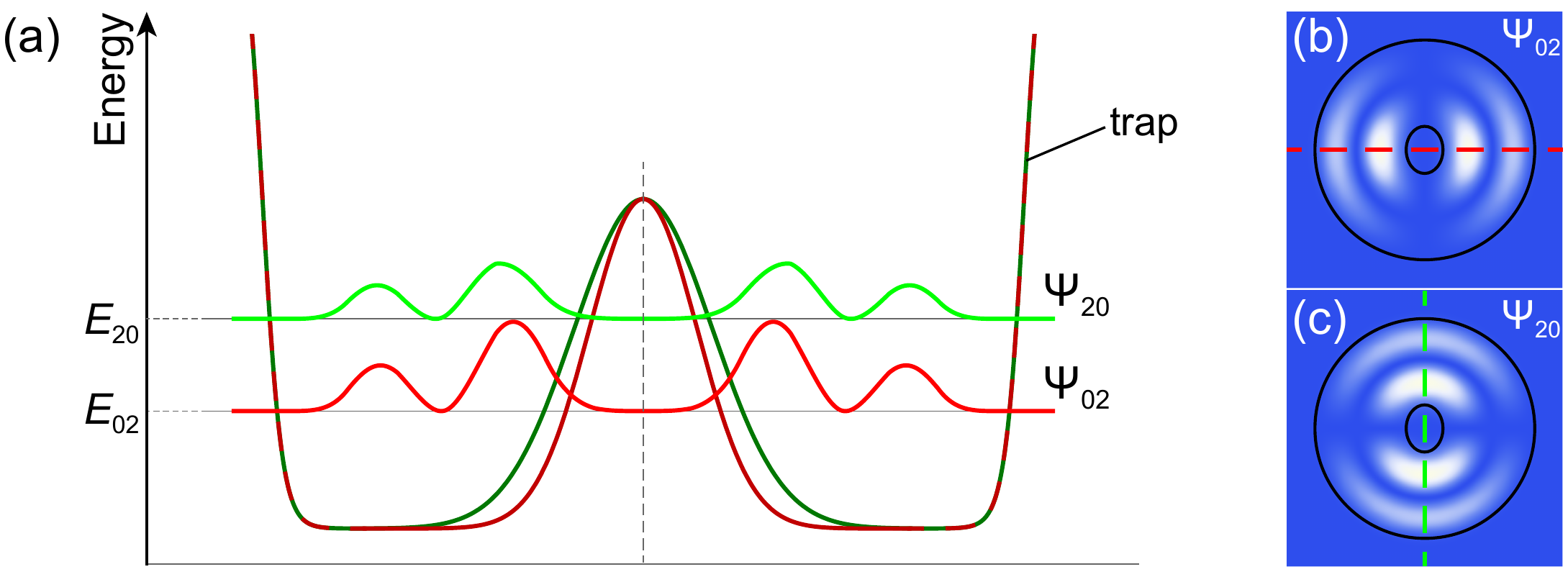}
\end{center}
\caption{ \label{FIG_OSC_Scheme}
Schematic of the splitting of eigenmodes in a trap with a reduced symmetry.
\clr{(a) The diagram shows the potential landscape of the trap, detailing the lifting of energy degeneracy of the $3p$-like modes. 
(b,c) Color maps showing $3p$-like polariton condensate modes, $\Psi _{02}$ and $\Psi _{20}$, in the trap.
The trap is formed by the stationary potential of the micropillar (outer edge) and by the cloud of excitons under the pump spot near the center of the micropillar (central peak).
The symmetry of the trap ring is reduced due to the elliptical shape of its central peak. 
Black contours on pales (b,c) give a hint about the shape of the trap.
Dark green (red) curve in panel (a) shows cross section of the trapping potential along the long (short) axis of the ellipse of the central peak. 
Light green (red) curve in (a) shows cross section of the polariton condensate eigenmode $\Psi_{20}$ ($\Psi_{02}$) quantized along the direction coinciding with the long (short) axis of the ellipse.
The corresponding cross sections are indicated by red and green dashed lines in panels (b) and (c).}
}
\end{figure}

\section*{Discussion}

The localization of a polariton condensate in an annular trap implies its spatial quantization in the radial direction.
In narrow ring traps, the radial modes are separated from each other by large distances in energy scale, and, as a rule, condensates occupy single ring modes characterized by the radial quantum number~$n=0$~\cite{PhysRevResearch3013072,PhysRevB97195149,PhysRevB103115309} \clr{(with no nodes in the radial direction)}.  
Increasing the width of the annulus of the trap results in exciting higher-order ring states with $n>0$~\cite{PNAS1118770,PhysRevB91045305,SedovNanosystPChM13608}.
The role of the radius of the (complex) trapping potential as the control parameter for selective excitation of different vortex states has been emphasized in Ref.~\cite{PhysRevB94134310} \clr{and discussed by us in the Introduction}.
Theoretical reasoning regarding the hierarchy of the states in the trap based on estimations of their condensation threshold pump powers is reported in Ref.~\cite{PhysRevB103115309}.
Here we discuss the \textit{fait accompli} observations of persistent azimuthal polariton current states in a polariton condensate having a form of two concentric rings.

\begin{figure}[tb!]
\begin{center}
\includegraphics[width=\linewidth]{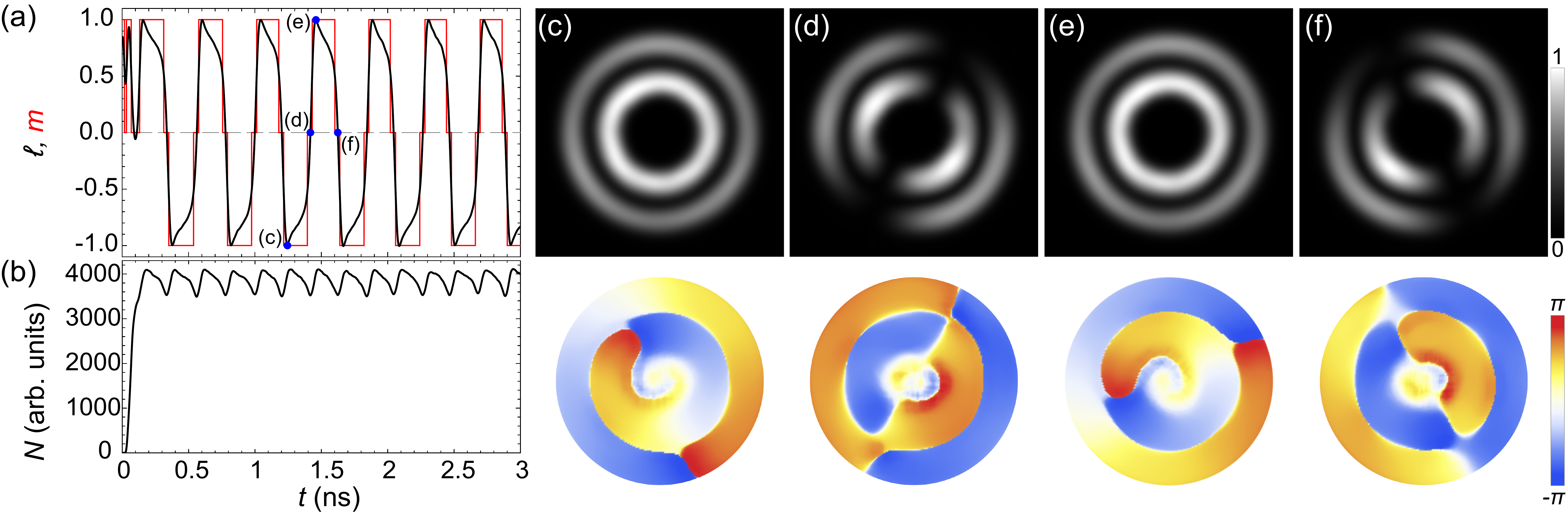}
\end{center}
\caption{ \label{FIG_SIMUL_Oscillations}
\clr{Simulation of evolution of the concentric ring polariton condensate in the regime of oscillating vorticity. 
Variation in time of OAM $\ell(t)$ (black curves) and the winding number $m(t)$ (red curves) (a), and occupation number of the polariton condensate~$N(t)$~(b).
Panels (c--f) show distribution in the micropillar plane of the density (top) and phase (bottom) of the polariton condensates at time moments indicated by blue dots in panel~(a), corresponding to $m = -1$ (c), $m = 0$ and $d_t \ell < 0$ (d), $m = 1$ (e), $m = 0$ and $d_t \ell > 0$ (f).}
}
\end{figure}

To support our experimental observations, we resort to numerical simulations.
For the simulations we use the generalized Gross-Pitaevskii equation for the polariton condensate wave function~$\Psi (t, \mathbf{r})$, coupled to the rate equation for the density of the reservoir of incoherent excitons, see Methods for details of the model.
Figure~\ref{FIG_SIMUL_Rings} shows the interferometry images and the phase distributions of the  simulated double concentric ring polariton condensates with the winding numbers of $m = 1$ and $m = 3$.
The double ring shape of the condensates in the simulations was achieved by adjusting the parameters responsible for the gain-loss balance in the micropillar plane, among which the most essential ones are the polariton losses, the stimulated scattering rate from the exciton reservoir to the condensate state and the width of the trap.
For inducing polariton azimuthal currents, we broke the rotational symmetry of the system by shifting the pump spot from the center of the pillar and supplementing the stationary potential with weak pertirbations.
This and other approaches for making the system chiral have been discussed by us in earlier papers~\cite{PhysRevB97195149,ACSPhoton71163,PhysRevResearch3013072} for single ring condensates.

Of the greatest interest among our observations is the experiment illustrated in Fig.~\ref{FIG_EXP_Superp}.
A double break of the circular interference fringes indicating a pair of compensating each other phase jumps could be inherent in a two-lobe (``dipole'') mode of a trap~\cite{PhysRevB82073303,PhysRevLett113200404,JETPL103313}.
However, the fact that the jumps coexist with the azimuthally homogeneous density distribution, make us reject this assumption as an option. 
A similar coexistence has been observed in a recent paper~\cite{CherbuninPaper2022}. 
The authors observe two phase jumps by $\pm \pi$ in two nearly diametrically opposed points of the phase ditribution extracted from the double-split interference fringes.
They interpret this as the result of ensemble averaging over multiple condensates stochastically changing their winding direction.
The stochastic nature of the change of $m$ in Ref.~\cite{CherbuninPaper2022} is possible due to the pulse-periodic regime of the non-resonant optical pumping, when each emerging condensate does not inherit vorticity of the condensate excited by the previous pulse. 

In our experiments, in contrast, the condensates are excited by the continuous wave pump, which eliminates the stochastic switching caused by stochastic initial conditions. 
We assume that we deal with the steady state regime of vorticity oscillating in time.
Averaging over the periodically replacing each other during oscillations vortex states of the condensate with oppositely directed currents could give  the desired interference image with doubly broken fringes. 

Among the options that could lead to oscillations of vorticity in time are excitation of limit cycles in the polariton system~\cite{PhysRevLett114193901,PhysRevB101085302} and beats of two close in energy eigenmodes of the trap~\cite{PhysRevB101115418}.
The former represent oscillating in time solutions that are the only stable solutions of a nonlinear problem in a considered range of parameters.
They are typically characterized by frequency combs in their photoluminescence spectrum~\cite{PhysRevLett114193901,PhysRevB101085302}.
This allows us to rule out the limit cycles as a possible interpretation of the observed oscillations.

As one has mentioned in Ref.~\cite{PhysRevResearch3033187,PhysRevB101115418},  oscillating solutions are fingerprints of superposition states, in particular, spatial modes of the trap.
Eigenmodes of the annular trap resemble eigenmodes of a pot-like trap in their shape except those with the density maximum in the center of the trap.
Let us focus on a couple of $3p$-like excited states, $(\Psi _{02} , \Psi _{20})$, which differ from the dipole ($2p$) modes by an additional circular node in the radial direction, see color maps in Figs.~\ref{FIG_OSC_Scheme}(b,c).
In the rotationally symmetric potential, the orthogonal modes are degenerate in energy.
Their linear superposition, $\Psi _{02} \pm i\Psi _{20}$, gives the double concentric ring condensate with the counterclockwise ($m=1$) or clockwise ($m=-1$) polariton currents.
Bringing the eigenmodes out from the degeneracy enables oscillations between them with frequency defined by their energy splitting, $E_{20} - E_{02}$.
Such splitting can be achieved by reducing the rotational symmetry of the trap to the reflection symmetry.
The authors of Ref.~\cite{PhysRevB97235303}, e. g., gave a small ellipticity to the optical trap.
The origin of such symmetry reduction is not clear in our experiment.
Nevertheless, in the annular  geometry, the shift of the pump spot or deformation of its shape may lead to the desired outcome.
In Fig.~\ref{FIG_OSC_Scheme}(a) we schematically show how the degeneracy of eigenmodes can be lifted by introducing the ellipticity of the central peak of the trapping potential.

\clr{The simulations of evolution of the polariton condensate in the regime of oscillating vorticity are shown in Fig.~\ref{FIG_SIMUL_Oscillations}.
Panels~\ref{FIG_SIMUL_Oscillations}(a) and~\ref{FIG_SIMUL_Oscillations}(b) show evolution in time of OAM $\ell(t)$, winding number $m(t)$ and occupancy of the condensate state $N(t)$.
For reducing the symmetry of the trap, we introduced ellipticity of the Gaussian pump spot~$s=0.87$ (see Methods for details of the pump).
For the initial conditions, the density distribution was taken in the form of random distribution, while the phase contained a seed in the form of a vortex with the winding number~$m=1$.
In the considered numerical experiment, both OAM $\ell$ and the winding number $m$ exhibits oscillations in the range from $-1$ to $+1$ with period estimated as about~$0.44$~ns, accompanied by oscillations of occupancy~$N$.
The period of oscillations can be varied in a considerable range, e.~g., by changing ellipticity of the pump spot~$s$ or the pump power, and it typically ranges from tenths to units of nanoseconds, which matches the estimations made in~Ref.~\cite{PhysRevResearch3013099}. 
Panels~\ref{FIG_SIMUL_Oscillations}(c)--\ref{FIG_SIMUL_Oscillations}(f) show the distribution in real space of the density and phase of the polariton condensate at different time moments (indicated in panel~\ref{FIG_SIMUL_Oscillations}(a)), corresponding to the winding number $m =\pm 1$ and $0$.
One can clearly see that at $m =\pm 1$ both the density and phase of the condensate meet the characteristics given above for the observed stationary states shown in Figs.~\ref{FIG_EXP_Single_and_Triple}(d--f,j--l).
The intermediate states with $m=0$ are the $3p$-like states representing superpositions of the current states.

\begin{figure}[tb!]
\begin{center}
\includegraphics[width=\linewidth]{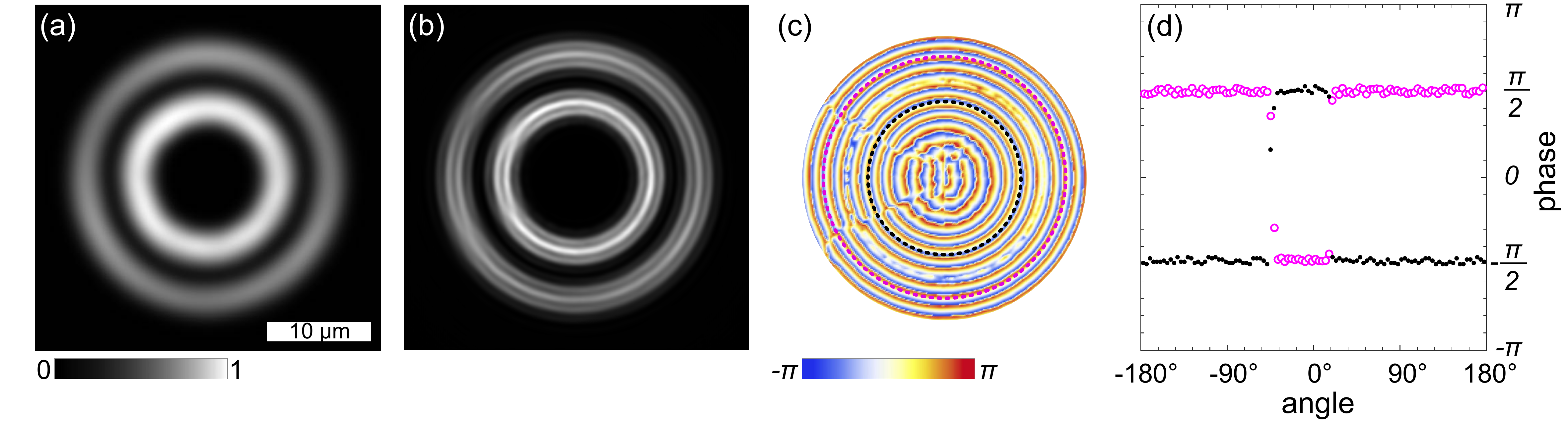}
\end{center}
\caption{ \label{FIG_SIMUL_Superp}
Simulation of the interferometry of the concentric ring polariton condensate with oscillating vorticity.
Meaning of panels is the same as in Fig.~\ref{FIG_EXP_Superp}.
\clr{The density distribution (a) and the interferometry image (b) were obtained by superimposing density distributions and interferometry images of two polariton condensates with $m = +1$ and~$-1$.
The relative phase distribution (c) was extracted from the interferometry image (b) using the Fourier-transform method adapted for closed-fringe patterns.}
}
\end{figure}

\begin{figure*}[tb!]
\begin{center}
\includegraphics[width=0.99\linewidth]{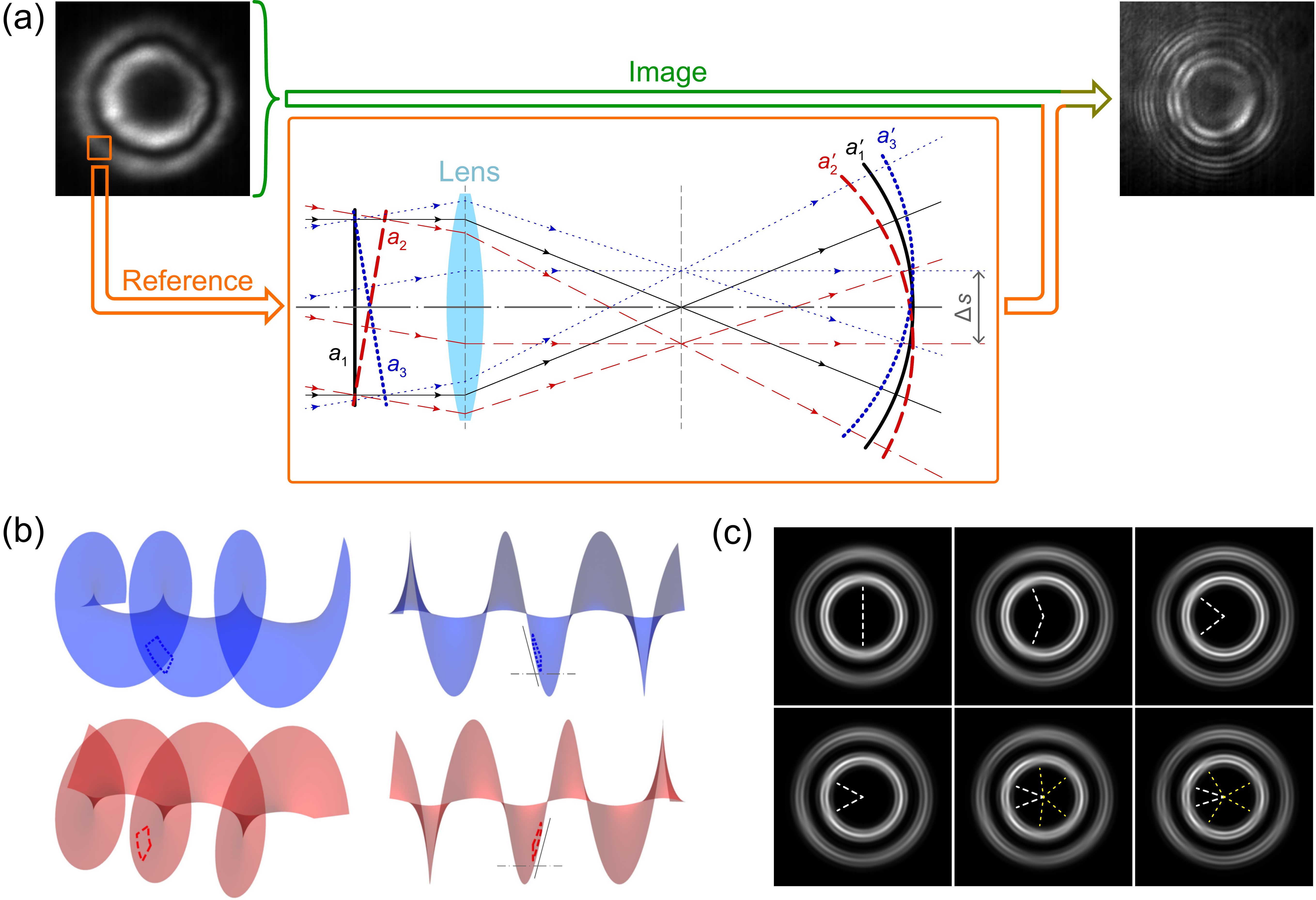}
\end{center}
\caption{ \label{FIG_EXP_Int_scheme}
\clr{Illustration to the formation of fringe breaks in the interference images for the polariton condensate in the regime of oscillating vorticity.
(a) Schematic of observation of interference of the condensate PL with a spherical reference wave.
The leftmost image is the PL density distribution of the condensate in the micropillar, the rightmost image is the interferometry image obtained from the interference of PL of the condensate with the reference spherical wave.
The  spherical wave is obtained by transforming a nearly plane wavefront of PL from a small peripheral region of the condensate using a focusing lens, see schematic in the solid orange frame.
Inclination of the plane wavefront of the incident wave results in the shift in the lens focal plane of the center of the outgoing spherical wave, cf. wavefronts $a_1$,~$a'_1$ and $a_3$,~$a'_3$ shown by red and blue dashed lines, respectively.
(b) Illustration of the inclination of the nearly flat wavefront, cut from the spiral wavefronts with opposite helicities. 
(c) Simulated examples of the averaged interferometry images with interference fringe breaks obtained at different shifts~$\Delta s$ between the centers of the reference spherical waves.
Each image represents a superposition of two interferometry images obtained from interference of the condensates of opposite vorticities and spherical waves correspondingly shifted in the image plane.
The shift $\Delta s$ in the panels increases from left to right and from top to bottom.
In the images, the shift is carried out in the vertical direction.
Positions of the fringe breaks are indicated by white dashed lines.
Yellow dashed lines indicate positions of additional fringe breaks that emerge at large~$\Delta s$.}
}
\end{figure*}

In Fig.~\ref{FIG_SIMUL_Superp} we numerically qualitatively reproduce the interferometry images emerging in the regime of oscillating vorticity, cf. Fig.~\ref{FIG_EXP_Superp}.
We superimpose the interferometry images obtained for the $m=+1$ and $m=-1$ states~(Fig.~\ref{FIG_SIMUL_Superp}(b)) and extract the phase from the resulting image~(Fig.~\ref{FIG_SIMUL_Superp}(c,d)).
The position of the fringe breaks is governed by two factors.
The first factor is the difference of the phases of the condensate modes contributing to the formation of the interference pattern, which determines the azimuthal position of the breaks. 
The change in the phase difference would result in the shift of the breaks in the azimuthal direction.
In the experiment, the phase difference remains constant, and it is determined by geometry of the trap and symmetry properties of the sample.

The second factor is the difference of the phases of the condensate PL and of the reference wave, which determines the relative position of the two fringe breaks.
The phase of the reference wave depends on the area of the condensate used for producing the reference beam.}
If the own phase distribution of the reference wave didn't change during observation of the oscillating steady state of the condensate, we would observe the breaks of the interference fringes in two diametrically opposed points.
However, the phase variation within the chosen area may affect the wavefront of the reference wave.

In brief, the wavefront of PL of the polariton condensate in the vortex state is helical. 
In a small area of the condensate, the phase variation of the wavefront is weak, so the front of the wave from this area is nearly flat but slightly tilted depending on the direction of vorticity.
The convex lens effectively converts the tilted flat wavefront to a spherical wavefront.
The center of the latter, however, turns out to be shifted from the optical axis of the lens.
This means that the position of the reference wave also oscillates simultaneously with steady state oscillations of the condensate vorticity.
Our simulations illustrated in Fig.~\ref{FIG_SIMUL_Superp} confirm that giving a small shift in the image plane between the spherical reference beams for the vortex states with $m=+1$ and $-1$ results in the convergence of the positions of the breaks on the interference fringes.
Herewith the real space PL distribution remains not anyhow affected by this procedure keeping the double concentric ring shape.
More details on the effect of the initial wavefront tilt on the reference wavefront see in Methods.

In conclusion, we have demonstrated the excitation of vortices in double concentric ring condensates of exciton polaritons in an effective annular trapping potential.
The concentric ring shape is subject to competition of gain and loss of the condensate in the radial direction, and it effectively depends on the width of the annulus.
In the considered geometry, the polariton condensates are characterized by similar vorticity in each condensate ring.
The interference of such states with the reference spherical wave gives interferometry images in the form of spirals.
We have also observed the polariton condensate with the steady state oscillations of vorticity in both rings.
On the averaged in time interferometry images, the oscillations result in even number of breaks of fringes, which position is determined by peculiarities of the reference wave.

\section*{Methods}

\subsection*{Sample and setup details}

The sample under our study represents a high quality (with Q-factor of about $1.6\cdot 10^4$) planar $5 \lambda /2 $ distributed Bragg reflector microcavity with an embedded set of quantum wells.
The width of the cavity layer varies along the microcavity plane providing the exciton-cavity photon detuning in the range of $-(0.5 \div 3.5)$~meV.
A set of cylindrical micropillars of different diameters is etched in the cavity plane.
In this manuscript, we study a micropillar of a diameter of $d = 30 \, \mu \text{m}$ with the exciton-photon detuning of~$-0.5$~meV.
In the experiment, the sample was kept in the helium-flow cryostat at a temperature of~4~K.

For exciting polariton condensates we use a cw Ti:sapphire laser.
The energy of the pump is tuned to the local minimum at the edge of the upper stop-band of the top Bragg mirror (about 110~meV above the bottom of the low polariton dispersion curve).
Full width at half maximum of the pump spot on the sample is about~$2\, \mu \text{m}$.

\subsection*{Numerical model}
For simulating persistent azimuthal polariton currents in concentric ring condensates we use the generalized Gross-Pitaevskii equation for the polariton wave function $\Psi (t,\mathbf{r})$~\cite{NJPhys14075020,PhysRevLett109216404,CherbuninPaper2022}:
\begin{equation}
\label{EqGPE}
i \hbar \partial _t \Psi (t,\mathbf{r}) = \left\{ [i \Lambda _0 n_{\text{R}} (t,\mathbf{r}) - 1] \frac{\hbar ^2}{2 M} \nabla ^2
 + V _{\text{eff}} (t,\mathbf{r}) 
+ \frac{i \hbar}{2} \left[ R n_{\text{R}} (t,\mathbf{r}) - \gamma \right] \right\} \Psi (t,\mathbf{r}),
\end{equation}
where $M$ is the effective mass of microcavity polaritons,
$V _{\text{eff}}  (t,\mathbf{r}) = V (r) + \alpha |\Psi (t,\mathbf{r})|^2 + \alpha _{\text{R}} n_{\text{R}} (t,\mathbf{r}) $ is the effective trapping potential composed of the stationary potential of the micropillar, $ V ({r})$, and the quasiparticle-interaction-induced potential.
$\alpha$ and $\alpha _{\text{R}}$ are the polariton-polariton interaction constant and the interaction constant of polaritons with the reservoir of incoherent excitons, respectively,
$n_{\text{R}} (t,\mathbf{r})$ is the density of the exciton reservoir.
The stationary potential is taken in the complex form, $V(r) = V_{\text{Re}} (r) + i V_{\text{Im}} (r)$.
The real part is taken as $V_{\text{Re}} (r) = V_0 \left\{ \tanh [a(r - d/2)] +1\right\}/2$ with $V_0$ and $d$ being the height of the potential and the diameter of the pillar, $a$ is the fitting parameter.
The imaginary part, $V_{\text{Im}} (r)$, is responsible for additional damping due to etching of the microcavity~\cite{PhysRevLett126075302,ResInOpt4100105}.
It can be obtained from $V_{\text{Re}} (r)$ by replacing ${V_0 \rightarrow V_0'}$, ${a \rightarrow a'}$ and ${d \rightarrow d'}$.

The rightmost imaginary term in Eq.~\eqref{EqGPE} is responsible for the gain-loss balance in the polariton condensate.
$R$ is the stimulated scattering rate from the incoherent exciton reservoir to the condensate,
$\gamma$ is the decay rate of polaritons.
The imaginary term in the kinetic energy in Eq.~\eqref{EqGPE} is responsible for the energy relaxation of expanding polaritons~\cite{NJPhys14075020,PhysRevLett109216404}.
$\Lambda _0$ is the energy relaxation constant, which we take as a fitting parameter.

Equation~\eqref{EqGPE} is coupled to the rate equation for the exciton reservoir density $n_{\text{R}} (t,\mathbf{r})$:
\begin{equation}
\label{EqResEq}
\partial _t n_{\text{R}} (t,\mathbf{r}) = P (\mathbf{r}) - \left[ \gamma _{\text{R}} + R |\Psi (t,\mathbf{r})|^2 \right] n_{\text{R}} (t,\mathbf{r}).
\end{equation}
The reservoir is excited by the optical pump of the Gaussian shape
\begin{equation}
P(\mathbf{r}) \propto \exp \left[ - \frac{(x - x_{\text{p}})^2}{2w^2}- \frac{(y - y_{\text{p}})^2}{2(sw)^2} \right]
\end{equation}
of width~$w$,
where the vector $(x_{\text{p}},y_{\text{p}})$ characterizes shift of the pump spot from the center of the pillar,
$s$~is responsible for the ellipticity of the pump spot.
$\gamma _{\text{R}}$ is the decay rate of the reservoir excitons.

For breaking the rotational symmetry of the problem, in our simulations we combine the slight shift and ellipticity of the pimp spot $P (\mathbf{r})$ with arbitrary initial conditions.

\subsection*{Values of the parameters}
The effective mass of polaritons is $M = 3 \cdot 10^{-5} m_{\text{e}}$, where $m_{\text{e}}$ is the free electron mass.
The polariton and exciton decay rates are taken as $\gamma = 0.025$~ps$^{-1}$ and $\gamma _{\text{R}} = 0.02$~ps$^{-1}$, respectively.
The stimulated scattering rate is taken as $\hbar R=0.1\, \text{meV} \, \mu \text{m}^2$.
The nonlinearity coefficients are taken as $\alpha = \alpha_{\text{R}}/2 = 3 \, \mu \text{eV} \, \mu \text{m}^2$.
The fitting parameter of the energy relaxation is $\Lambda _0 = 0.01 \, \mu \text{m}^2$.
The pump width is $w = 2.5 \, \mu\text{m}$.
The parameters of the stationary potential: $V_0 = - 3 V_0' = 3 \, \text{meV}$, $a = a' = 4 \, \mu\text{m}^{-1}$, $d = 30 \, \mu\text{m}$ and $d ' = 30.8\, \mu\text{m}$.

\subsection*{Details of obtaining a spherical reference wave}

In the interferometry measurements, we used the following method for obtaining a coherent spherical reference wave of the same frequency as PL of the condensate.
At the periphery of the observation zone, we isolated a small area of the condensate.
Then we passed the luminescence signal from this area through a converging lens, which is separated from the image plane by a distance much greater than its focal length. 
The converging lens converts a plain wavefront of the incident radiation to a spherical wavefront, which converges beyond the lens focal length, see schematic in Fig.~\ref{FIG_EXP_Int_scheme}(a).

When the polariton condensate contains a vortex, its PL is characterized by a helical wavefront.
When we cut a small area from the condensate, the wavefront within this area is close to plain, but it still keeps some deviations, in particular, inclination with respect to the optical axis of the lens, see Fig.~\ref{FIG_EXP_Int_scheme}(b).
The flatness of the wavefront is smaller, the larger the cut area. 
Remarkably, the inclination of the wavefronts of two opposite vortices in the same area of the condensate is opposite, compare blue and red images in Fig.~\ref{FIG_EXP_Int_scheme}(b).

The inclined wavefront is still converted by the converging lens to a spherical wavefront, however the center of this  wave is shifted in a focal plane from the lens principal focus by some distance depending on the inclination~\cite{KrugerBook1963}, cf. black (solid), blue (dotted) and red (dashed) curves in Fig.~\ref{FIG_EXP_Int_scheme}(a).
Obviously, for the oppositely inclined plane wavefronts, the centers of the spherical wavefronts are shifted in opposite directions.

Written above allows to conclude that for the oppositely winding polariton vortex states emerging in similar experimental conditions, the reference spherical waves obtained from the same area of the condensate are weakly shifted with respect to each other by some distance $\Delta s$, which results in the corresponding shift of the interference fringe breaks.
\clr{Figure~\ref{FIG_EXP_Int_scheme}(c) shows simulated examples of the averaged interferometry images obtained at different shifts~$\Delta s$ between the centers of the reference spherical wavefronts.
Rapprochement of the fringe breaks with increasing $\Delta s $ is clearly seen.}

\section*{Acknowledgements}

This work was done under the support of the Saint Petersburg State University (Grant No. 94030557).
A.K. and P.S. acknowledge the support of Westlake University, Project 041020100118 and Program 2018R01002 funded by Leading Innovative and Entrepreneur Team Introduction Program of Zhejiang Province of China.
A.K.~acknowledges support from the Moscow Institute of Physics and Technology under the Priority 2030 Strategic Academic Leadership Program.
Work of V.L., V.K., and A.K. was partially supported the RFBR (Grant No. 15-52-12032).
Numerical simulations were carried out within the state assignment in the field of scientific activity of the Ministry of Science and Higher Education of the Russian Federation (theme FZUN-2020-0013, state assignment of VlSU).

\bibliographystyle{unsrt}
\bibliography{doubleRingBibl}

\begin{thebibliography}{10}

\bibitem{kavokinBook2017}
A.~Kavokin, J.J. Baumberg, G.~Malpuech, and F.P. Laussy.
\newblock {\em Microcavities}.
\newblock Series on Semiconductor Science and Technology. OUP Oxford, 2
  edition, 2017.

\bibitem{ScienceScience199}
Hui Deng, Gregor Weihs, Charles Santori, Jacqueline Bloch, and Yoshihisa
  Yamamoto.
\newblock Condensation of semiconductor microcavity exciton polaritons.
\newblock {\em Science}, 298(5591):199--202, 2002.

\bibitem{NatPhys10803}
Tim Byrnes, Na~Young Kim, and Yoshihisa Yamamoto.
\newblock Exciton--polariton condensates.
\newblock {\em Nature Physics}, 10(11):803--813, 2014.

\bibitem{PhysRevX5021025}
A.~Metelmann and A.~A. Clerk.
\newblock Nonreciprocal photon transmission and amplification via reservoir
  engineering.
\newblock {\em Phys. Rev. X}, 5:021025, Jun 2015.

\bibitem{PhysRevA98053812}
Maximilian Keck, Davide Rossini, and Rosario Fazio.
\newblock Persistent currents by reservoir engineering.
\newblock {\em Phys. Rev. A}, 98:053812, Nov 2018.

\bibitem{PhysRevB85155320}
M.~A\ss{}mann, F.~Veit, M.~Bayer, A.~L\"offler, S.~H\"ofling, M.~Kamp, and
  A.~Forchel.
\newblock All-optical control of quantized momenta on a polariton staircase.
\newblock {\em Phys. Rev. B}, 85:155320, Apr 2012.

\bibitem{PhysRevB102201114}
Nikita Stroev and Natalia~G. Berloff.
\newblock Managing the flow of liquid light.
\newblock {\em Phys. Rev. B}, 102:201114, Nov 2020.

\bibitem{PhysRevResearch3013099}
Yan Xue, Igor Chestnov, Evgeny Sedov, Evgeniy Kiktenko, Aleksey~K. Fedorov,
  Stefan Schumacher, Xuekai Ma, and Alexey Kavokin.
\newblock Split-ring polariton condensates as macroscopic two-level quantum
  systems.
\newblock {\em Phys. Rev. Research}, 3:013099, Jan 2021.

\bibitem{NatPhys6527}
D.~Sanvitto, F.~M. Marchetti, M.~H. Szyma{\'n}ska, G.~Tosi, M.~Baudisch, F.~P.
  Laussy, D.~N. Krizhanovskii, M.~S. Skolnick, L.~Marrucci, A.~Lema{\^\i}tre,
  J.~Bloch, C.~Tejedor, and L.~Vi{\~n}a.
\newblock Persistent currents and quantized vortices in a polariton superfluid.
\newblock {\em Nature Physics}, 6(7):527--533, 2010.

\bibitem{RevModPhys85299}
Iacopo Carusotto and Cristiano Ciuti.
\newblock Quantum fluids of light.
\newblock {\em Rev. Mod. Phys.}, 85:299--366, Feb 2013.

\bibitem{PhysRevB97195149}
V.~A. Lukoshkin, V.~K. Kalevich, M.~M. Afanasiev, K.~V. Kavokin,
  Z.~Hatzopoulos, P.~G. Savvidis, E.~S. Sedov, and A.~V. Kavokin.
\newblock Persistent circular currents of exciton-polaritons in cylindrical
  pillar microcavities.
\newblock {\em Phys. Rev. B}, 97:195149, May 2018.

\bibitem{ACSPhoton71163}
Evgeny Sedov, Vladimir Lukoshkin, Vladimir Kalevich, Zacharias Hatzopoulos,
  Pavlos Savvidis, and Alexey Kavokin.
\newblock Persistent currents in half-moon polariton condensates.
\newblock {\em ACS Photonics}, 7(5):1163--1170, 05 2020.

\bibitem{PhysRevResearch3013072}
E.~S. Sedov, V.~A. Lukoshkin, V.~K. Kalevich, P.~G. Savvidis, and A.~V.
  Kavokin.
\newblock Circular polariton currents with integer and fractional orbital
  angular momenta.
\newblock {\em Phys. Rev. Research}, 3:013072, Jan 2021.

\bibitem{SciRep1122382}
Evgeny Sedov, Sergey Arakelian, and Alexey Kavokin.
\newblock Spontaneous symmetry breaking in persistent currents of spinor
  polaritons.
\newblock {\em Scientific Reports}, 11(1):22382, 2021.

\bibitem{PhysRevB101104308}
A.~V. Yulin, A.~V. Nalitov, and I.~A. Shelykh.
\newblock Spinning polariton vortices with magnetic field.
\newblock {\em Phys. Rev. B}, 101:104308, Mar 2020.

\bibitem{NatPhysRev4435}
Alexey Kavokin, Timothy C.~H. Liew, Christian Schneider, Pavlos~G. Lagoudakis,
  Sebastian Klembt, and Sven Hoefling.
\newblock Polariton condensates for classical and quantum computing.
\newblock {\em Nature Reviews Physics}, 4(7):435--451, 2022.

\bibitem{PhysRevB103075305}
Franziska Barkhausen, Matthias Pukrop, Stefan Schumacher, and Xuekai Ma.
\newblock Structuring coflowing and counterflowing currents of polariton
  condensates in concentric ring-shaped and elliptical potentials.
\newblock {\em Phys. Rev. B}, 103:075305, Feb 2021.

\bibitem{PhysRevB104165305}
I.~Chestnov, A.~Yulin, I.~A. Shelykh, and A.~Kavokin.
\newblock Dissipative josephson vortices in annular polariton fluids.
\newblock {\em Phys. Rev. B}, 104:165305, Oct 2021.

\bibitem{PhysRevB100245304}
S.~Mukherjee, D.~M. Myers, R.~G. Lena, B.~Ozden, J.~Beaumariage, Z.~Sun,
  M.~Steger, L.~N. Pfeiffer, K.~West, A.~J. Daley, and D.~W. Snoke.
\newblock Observation of nonequilibrium motion and equilibration in polariton
  rings.
\newblock {\em Phys. Rev. B}, 100:245304, Dec 2019.

\bibitem{YaoZBpaper2022}
Q.~Yao, E.~Sedov, S.~Mukherjee, J.~Beaumariage, B.~Ozden, K.~West, L.~Pfeiffer,
  A.~Kavokin, and D.~W. Snoke.
\newblock Ballistic transport of a polariton ring condensate with spin
  precession.
\newblock {\em Phys. Rev. B}, 106:245309, Dec 2022.

\bibitem{PNAS1118770}
Alexander Dreismann, Peter Cristofolini, Ryan Balili, Gabriel Christmann,
  Florian Pinsker, Natasha~G. Berloff, Zacharias Hatzopoulos, Pavlos~G.
  Savvidis, and Jeremy~J. Baumberg.
\newblock Coupled counterrotating polariton condensates in optically defined
  annular potentials.
\newblock {\em Proceedings of the National Academy of Sciences},
  111(24):8770--8775, 2014.

\bibitem{PhysRevB92035305}
A.~Askitopoulos, T.~C.~H. Liew, H.~Ohadi, Z.~Hatzopoulos, P.~G. Savvidis, and
  P.~G. Lagoudakis.
\newblock Robust platform for engineering pure-quantum-state transitions in
  polariton condensates.
\newblock {\em Phys. Rev. B}, 92:035305, Jul 2015.

\bibitem{PhysRevB103115309}
E.~D. Cherotchenko, H.~Sigurdsson, A.~Askitopoulos, and A.~V. Nalitov.
\newblock Optically controlled polariton condensate molecules.
\newblock {\em Phys. Rev. B}, 103:115309, Mar 2021.

\bibitem{PhysRevB97045303}
Yongbao Sun, Yoseob Yoon, Saeed Khan, Li~Ge, Mark Steger, Loren~N. Pfeiffer,
  Ken West, Hakan~E. T\"ureci, David~W. Snoke, and Keith~A. Nelson.
\newblock Stable switching among high-order modes in polariton condensates.
\newblock {\em Phys. Rev. B}, 97:045303, Jan 2018.

\bibitem{PhysRevB107045302}
Ekaterina Aladinskaia, Roman Cherbunin, Evgeny Sedov, Alexey Liubomirov, Kirill
  Kavokin, Evgeny Khramtsov, Mikhail Petrov, P.~G. Savvidis, and Alexey
  Kavokin.
\newblock Spatial quantization of exciton-polariton condensates in optically
  induced traps.
\newblock {\em Phys. Rev. B}, 107:045302, Jan 2023.

\bibitem{SedovNanosystPChM13608}
E.~S. Sedov, V.~A. Lukoshkin, V.~K. Kalevich, I.~Yu. Chestnov, Z.~Hatzopoulos,
  P.~G. Savvidis, and A.~V. Kavokin.
\newblock Double ring polariton condensates with polariton vortices.
\newblock {\em Nanosystems: Phys. Chem. Math.}, 13:608–614, Dec 2022.

\bibitem{PhysRevB91045305}
V.~K. Kalevich, M.~M. Afanasiev, V.~A. Lukoshkin, D.~D. Solnyshkov,
  G.~Malpuech, K.~V. Kavokin, S.~I. Tsintzos, Z.~Hatzopoulos, P.~G. Savvidis,
  and A.~V. Kavokin.
\newblock Controllable structuring of exciton-polariton condensates in
  cylindrical pillar microcavities.
\newblock {\em Phys. Rev. B}, 91:045305, Jan 2015.

\bibitem{APL92042119}
F.~Stokker-Cheregi, A.~Vinattieri, F.~Semond, M.~Leroux, I.~R. Sellers,
  J.~Massies, D.~Solnyshkov, G.~Malpuech, M.~Colocci, and M.~Gurioli.
\newblock Polariton relaxation bottleneck and its thermal suppression in bulk
  $\text{GaN}$ microcavities.
\newblock {\em Applied Physics Letters}, 92(4):042119, 2008.

\bibitem{PhysRevB65205310}
R.~Butt\'e, G.~Delalleau, A.~I. Tartakovskii, M.~S. Skolnick, V.~N. Astratov,
  J.~J. Baumberg, G.~Malpuech, A.~Di~Carlo, A.~V. Kavokin, and J.~S. Roberts.
\newblock Transition from strong to weak coupling and the onset of lasing in
  semiconductor microcavities.
\newblock {\em Phys. Rev. B}, 65:205310, Apr 2002.

\bibitem{SciRep75542}
R.~Jayaprakash, F.~G. Kalaitzakis, G.~Christmann, K.~Tsagaraki, M.~Hocevar,
  B.~Gayral, E.~Monroy, and N.~T. Pelekanos.
\newblock Ultra-low threshold polariton lasing at room temperature in a
  $\text{GaN}$ membrane microcavity with a zero-dimensional trap.
\newblock {\em Scientific Reports}, 7(1):5542, 2017.

\bibitem{PhysRevB94134310}
Alexey~V. Yulin, Anton~S. Desyatnikov, and Elena~A. Ostrovskaya.
\newblock Spontaneous formation and synchronization of vortex modes in
  optically induced traps for exciton-polariton condensates.
\newblock {\em Phys. Rev. B}, 94:134310, Oct 2016.

\bibitem{PhysRevB82073303}
Ga\"el Nardin, Konstantinos~G. Lagoudakis, Barbara Pietka, Fran\ifmmode
  \mbox{\c{c}}\else~\c{c}\fi{}ois Morier-Genoud, Yoan L\'eger, and Beno\^{\i}t
  Deveaud-Pl\'edran.
\newblock Selective photoexcitation of confined exciton-polariton vortices.
\newblock {\em Phys. Rev. B}, 82:073303, Aug 2010.

\bibitem{PhysRevLett113200404}
Robert Dall, Michael~D. Fraser, Anton~S. Desyatnikov, Guangyao Li, Sebastian
  Brodbeck, Martin Kamp, Christian Schneider, Sven H\"ofling, and Elena~A.
  Ostrovskaya.
\newblock Creation of orbital angular momentum states with chiral polaritonic
  lenses.
\newblock {\em Phys. Rev. Lett.}, 113:200404, Nov 2014.

\bibitem{JETPL103313}
V.~A. Lukoshkin, V.~K. Kalevich, M.~M. Afanasiev, K.~V. Kavokin, S.~I.
  Tsintzos, P.~G. Savvidis, Z.~Hatzopoulos, and A.~V. Kavokin.
\newblock Controlled switching between quantum states in the exciton--polariton
  condensate.
\newblock {\em JETP Letters}, 103(5):313--315, 2016.

\bibitem{CherbuninPaper2022}
J.~Borat, Roman Cherbunin, Evgeny Sedov, Ekaterina Aladinskaia, Alexey
  Liubomirov, Valentina Litvyak, Mikhail Petrov, Xiaoqing Zhou, Z.~Hatzopoulos,
  Alexey Kavokin, and P.~G. Savvidis.
\newblock Stochastic circular persistent currents of exciton polaritons, 2022.

\bibitem{PhysRevLett114193901}
K.~Rayanov, B.~L. Altshuler, Y.~G. Rubo, and S.~Flach.
\newblock Frequency combs with weakly lasing exciton-polariton condensates.
\newblock {\em Phys. Rev. Lett.}, 114:193901, May 2015.

\bibitem{PhysRevB101085302}
Seonghoon Kim, Yuri~G. Rubo, Timothy C.~H. Liew, Sebastian Brodbeck, Christian
  Schneider, Sven H\"ofling, and Hui Deng.
\newblock Emergence of microfrequency comb via limit cycles in dissipatively
  coupled condensates.
\newblock {\em Phys. Rev. B}, 101:085302, Feb 2020.

\bibitem{PhysRevB101115418}
C.~Leblanc, G.~Malpuech, and D.~D. Solnyshkov.
\newblock High-frequency exciton-polariton clock generator.
\newblock {\em Phys. Rev. B}, 101:115418, Mar 2020.

\bibitem{PhysRevResearch3033187}
I.~Chestnov, Y.~G. Rubo, A.~Nalitov, and A.~Kavokin.
\newblock Pseudoconservative dynamics of coupled polariton condensates.
\newblock {\em Phys. Rev. Research}, 3:033187, Aug 2021.

\bibitem{PhysRevB97235303}
A.~Askitopoulos, A.~V. Nalitov, E.~S. Sedov, L.~Pickup, E.~D. Cherotchenko,
  Z.~Hatzopoulos, P.~G. Savvidis, A.~V. Kavokin, and P.~G. Lagoudakis.
\newblock All-optical quantum fluid spin beam splitter.
\newblock {\em Phys. Rev. B}, 97:235303, Jun 2018.

\bibitem{NJPhys14075020}
Michiel Wouters.
\newblock Energy relaxation in the mean-field description of polariton
  condensates.
\newblock {\em New Journal of Physics}, 14(7):075020, jul 2012.

\bibitem{PhysRevLett109216404}
E.~Wertz, A.~Amo, D.~D. Solnyshkov, L.~Ferrier, T.~C.~H. Liew, D.~Sanvitto,
  P.~Senellart, I.~Sagnes, A.~Lema\^{\i}tre, A.~V. Kavokin, G.~Malpuech, and
  J.~Bloch.
\newblock Propagation and amplification dynamics of 1d polariton condensates.
\newblock {\em Phys. Rev. Lett.}, 109:216404, Nov 2012.

\bibitem{PhysRevLett126075302}
J.~Beierlein, E.~Rozas, O.~A. Egorov, M.~Klaas, A.~Yulin, H.~Suchomel, T.~H.
  Harder, M.~Emmerling, M.~D. Mart\'{\i}n, I.~A. Shelykh, C.~Schneider,
  U.~Peschel, L.~Vi\~na, S.~H\"ofling, and S.~Klembt.
\newblock Propagative oscillations in codirectional polariton waveguide
  couplers.
\newblock {\em Phys. Rev. Lett.}, 126:075302, Feb 2021.

\bibitem{ResInOpt4100105}
Irina Sedova and Evgeny Sedov.
\newblock Polarization conversion in a polariton three-waveguide coupler.
\newblock {\em Results in Optics}, 4:100105, 2021.

\bibitem{KrugerBook1963}
M.~Ya. Kruger, V.~A. Panov, V.~V. Kulagin, G.~V. Pogarev, Ya.~M. Kruger, and
  A.~M. Levinzon.
\newblock {\em Handbook of the designer of optical-mechanical devices}.
\newblock MASHGIZ, Moscow, Leningrad, 1963.

\end{thebibliography}

\end{document}